# A biomechanical study of neck strength and impact dynamics on head and neck injury parameters

Rahid Zaman, Muhammad Ibrahim Hossain, Ahmed Zubayer Raiyan, Shuvo Chowdhury, Aaron Jackson, Arthur Thomas Koster, Ashfaq Adnan

## Abstract

Traumatic brain injuries (TBI) are considered a silent epidemic. It affects many people, from automobiles to sports to service members. In this study, we employed a musculoskeletal head-neck model to understand the effect of impact locations, characteristics, and neck strength on head and neck injury severity. Three types of impact forces were studied: low-velocity impact (LVI), intermediate-velocity impact (IVI), and high-velocity impact (HVI). We investigated six parameters: linear and rotational accelerations, the Generalized Acceleration Model For Brain Injury Threshold (GAMBIT), neck force, neck moment, and Neck Injury Criteria (NIC). We consider seven impact locations, three neck strengths, and three impact characteristics. We studied a total of 63 cases. It was found that the linear accelerations do not change much with different neck strengths and impact locations. The impact locations have a significant effect on head and neck injury parameters, and anterolateral impact is the most risky impact location for both head and neck. Rotational acceleration is considered to cause mild TBI (mTBI) and diffuse axonal injuries. The maximum average rotational acceleration is for anterolateral eccentric impact (4176 $rad/s^2$) which is 4.75 times more than the average anterior central impact (879 $rad/s^2$). The lateral impacts generate about 10% more linear accelerations than anterior and posterior impacts. Neck forces do not vary more than 20% with impact locations and neck strength. The average head and neck injury parameters do not vary more than 10% based on neck strength. Impact characteristics have a significant role in GAMBIT and NIC. The average GAMBIT for IVI and HVI were 1.44 and 1.54 times higher than LVI. In summary, the anterolateral eccentric impact has a higher probability of head and neck injury than the other six impact locations. These findings provide objective evidence that can inform injury prevention strategies as well as aid tissue and cellular level studies.

## 1 Introduction

An estimated 1.6 to 3.8 million sport and recreation-related traumatic brain injuries (TBIs) occur in the USA annually (Borich et al., 2013). Over 6 million passenger car accidents every year, and more than 42,514 people lost their lives in 2022 in the USA (Insurance Institute for Highway Safety and Highway Loss Data Institute (IIHS-HLDI), 2024). According to CDC, only in 2019, 60,611 TBI-related deaths occurred in the USA (Centers for Disease Control and Prevention (CDC), 2019). A mild TBI (mTBI), also known as concussion, has been widely underreported. Even in fully equipped settings, like sports, 66% of sports-related TBI go unreported (McCrea et al., 2004). According to the Defense Health Agency, 505,896 service members faced mild to severe TBI injuries from 2000 to 2024 (Defense Health Agency (DHA), 2024). In a study, it was found that 57% of service members who face mTBI did not seek medical care. (Escolas et al., 2020). So, The percentage of the civil population to ignores mTBI is really a concern. That's' why TBI is frequently referred to as a "silent epidemic" (Takhounts et al., 2011). TBI is a complex pathophysiological process induced by mechanical loading of the brain. However, the complications from TBI has long-term effects and can alter a person's thinking, sensation, language, or emotions, which may not be immediately apparent. Most of the life-threating injuries from contact sports, automobiles, and blasts involve torso, spine, and head. Therefore, it is no surprise that very detailed models of the human head and neck is a vital tool to assess any injury.

The cervical spine is a pivotal structure of head dynamics and controls the head response after external loading. It establishes the mechanical linkage between the torso and head and stabilizes the head using cervical disc and neck muscle tissue resistances. Since all anthropomorphic test devices (ATD) use pin joints to connect the vertebrae, the intervertebral rotation cannot be mimicked with high biofidelity. Therefore, the evaluation of the rotational acceleration-based head injury criterion and Neck Injury Criterion (NIC) is not reliable with ATD experiments. Also, the ATDs (Hybrid III, BioRID, etc.) are not scalable enough to make them subject-specific. It works based on percentile. That's why there is a growing interest in developing data using high biofidelity computational models. At the initial stages, lumped parameter models and multibody models of head and neck were developed. Most of those models are based on simplified rigid linkage. With the increase in computational efficiency, finite element models (FEM) became popular. Finite element models were widely used to study the head and neck motion for whiplash. However, it is very difficult for the FEM model to exactly mimic the nature of muscle tissues and their interactions with the surrounding anatomical structures, such as muscle wrapping, dynamics, muscle excitations and contractions, and interactions with tendons and skeletons.

However, to understand the response of the head and neck under external loadings, it is essential to integrate all these properties into the model to estimate a model's muscle force-generating capacity. The reliability of muscle force-generating capacity is a contributing factor in determining the head-neck model response. Not only does the shape and alignment of the cervical spine stabilize the head-neck joints, but the ligament and voluntarily controlled muscles also play a contributing factor (Fanta et al., 2013). Computational musculoskeletal models can be a very effective tool for getting high biofidelity data as they contain muscle dynamics, wrapping, and joint kinematics data. Computational musculoskeletal models are scalable to subject-specific and can provide segmental acceleration, velocity, and force information for a specific subject. It can help us to study the cause-and-effect of various muscle and joint physiological properties of head and neck injuries. Furthermore, computational musculoskeletal models take less time and are less expensive than cadaveric or dummy tests. However, the musculoskeletal models need to be validated to ensure that they can indeed predict the experimental results. Once validated, they can be extended to predict results that are either difficult or impossible to generate experimentally. Musculoskeletal models have been widely used for low-speed human movements like weight lifting, walking, running etc. Recently, there have been some studies where the musculoskeletal model has been used for intermediate velocity impacts such as NFL (Jin et al., 2017; Kuo et al., 2019; Mortensen et al., 2018, 2020) and Rugby impacts (Silvestros et al., 2024). However, head and neck injuries because of low-velocity and high-velocity impacts were not studied much using computational musculoskeletal models.

The effect of neck strength on the concussions is still under investigation. Most of the studies show that stronger and stiffer necks reduce the risk of concussion (Eckner et al., 2014; Fanta et al., 2013; Hasegawa et al., 2014; Mang et al., 2015). However, some studies disagree with this (Mihalik et al., 2011; Schmidt et al., 2014). In a study, the effect of anterior central, posterior central, lateral central, and posterolateral impacts are analyzed for sports-related concussive and sub-concussive impacts (Mortensen et al., 2020). They found that neck strenth has statistical significance on head injury criteria. In another study, they consider cranial anterior, cranial posterior, lateral mid-posterior, material mid-anterior, and lateral interior impacts from rugby players. However, they studied only neck injuries. Anterior eccentric and posterior eccentric impacts were not studied for both head and neck injuries. Also, in all those studies, sports impacts were only the source of input data.

Schmidt et al. (Schmidt et al., 2014) also studied multiple football players and mentioned in their study that greater cervical stiffness reduced the odds of sustaining higher-magnitude head impacts. However, they did not find any strong correlation between stronger and larger neck muscles and the mitigation of head impact injury. In this study (Schmidt et al., 2014), it is also mentioned that one possible reason may be the risk compensation phenomenon. That means football players with stronger and larger cervical musculature may be confident that they are more protected from head and neck injuries and may engage in more violent collisions. On the other hand, weaker subjects tend to avoid violent collisions. Furthermore, it was also reported that without visual perception, muscle activation takes 0.027s after an impact, while with visual perception, muscle activation takes 0.127s before the impact (Fanta et al., 2013). Also, it is reported that head injury criteria show about 30% decrease with visual perception. For these reasons, the studies and decisions about the impact of neck strength on head injuries, based only on sports data, may mislead the findings.

To fill up these gaps, in our study, we consider three types of impact data: automobile as low-velocity impact (LVI), NFL as intermediate velocity impact (IVI), and blast as high-velocity impact (HVI). Thus, we make our data independent of visual perception. Also, we kept the muscle activation level the same for all cases. We consider three neck strength: low force capacity (sedentary or untrained), mid force capacity (recreationally active), and High force capacity (Athletic). For all these impacts characteristics and neck strengths, we consider seven impact locations: anterior concentric, anterior eccentric, posterior concentric, posterior eccentric, lateral concentric, posterolateral eccentric, and anterolateral eccentric. In total, we consider 63 cases, to effect of impact locations, impact characteristics, and neck strength on headn and neck injuries.

## 2  Method:

### 2.1  Head-neck dynamics

The dynamics of the human musculoskeletal system can be formulated with the Euler-Lagrange equations (Zuo et al., 2024) as per Eq. (1):

$$M(q)\ddot{q} + c(q,\dot{q}) = J_m^T f_m(act) + J_c^T f_c + \tau_{ext} \qquad (1)$$

Where, $q$, $\dot{q}$, and $\ddot{q}$ are the positions, velocities, and acceleration of the joints, $M(q)$ is the mass distribution matrix which contains masses and inertial properties of the body segments, $c(q,\dot{q})$ is the Coriolis and centrifugal force vector which arises when Newton's laws of motion are applied in reference frames that are fixed to rotating bodies, $f_m(act)$ is the vector representing muscle forces generated by all muscle-tendon units, and is determined by muscle activations $(act)$, $f_c$ is the constraint force, and $\tau_{ext}$ is the external toque from the interaction with environments forces e.g., ground reaction forces, impact forces, $J_m$ and $J_c$ are Jacobian matrices that map forces to the space of generalized coordinates.

The acceleration of the body in response to muscle forces and other loads can be computed using the equations of motion for the body as per Eq. (2)

$$\ddot{q} = M^{-1}(q)\{-c(q,\dot{q}) + J_m^T f_m(act) + J_c^T f_c + \tau_{ext}\} \qquad (2)$$

The process to get the muscle force vectors is shown in Eq. (4).

### 2.2  Muscle-force generating parameters

The relationship between muscle activation $a(t)$ and muscle excitation $u(t)$ can be expressed as a first-order ordinary differential equation:

$$\dot{a} = \frac{u(t) - a(t)}{\tau(a,u)} \qquad (3)$$

$$\text{where } \tau(a,u) = \begin{cases} \tau_A(0.5 + 1.5a(t)), & \text{if } u(t) > a(t) \\ \dfrac{\tau_D}{0.5 + 1.5a(t)}, & \text{otherwise} \end{cases}$$

$\tau_A$ and $\tau_D$ represent the activation and deactivation time constants. $\tau_A$ is smaller that $\tau_D$. Although the values varies based on age, muscle, composition and other factors, typical values are 10 and 40 ms, respectively (Uchida & Delp, 2021). This relationship shows that the rate fo activation slows as the activation level increases due to less amount of calcium release and diffusion. Similarly, the rate of deactivation slows as the muscle activation level decreases. The force-length-velocity relationship of a muscle largely depends on the muscle activation dynamics.

Muscle forces can be computed from the activation $a(t)$, normalized fiber length ($l^M$), and normalized muscle velocity ($v^M$). Muscle force at particular muscle length, velocity and activation is the product of the corresponding values on the force-length and force-velocity curves.

$$f^M(t) = f_o^M[a(t)\, f^L(l^M(t))\, f^v(v^M(t)) + f^{PE}(l^M(t))] \qquad (4)$$

Where $f^M(t)$, $f_o^M$, and $a$ are fiber force, maximum isometric force, and muscle activation, respectively.

The total generated muscle force is a function of muscle fiber length and, muscle velocity. The force-length curve, $f^L(l^M(t))$ Provides the relationship between muscle length and generated force at a specific time. The force-velocity curve $f^v(v^M(t))$ provides the relationship between muscle velocity and generated force at a specific time. As our impact characteristics are classified based on velocity and duration, we considered neck -strength to be a function of the force-velocity curve $f^v(v^M(t))$. In musculoskeletal modeling, maximum force output can be expressed as Eccentric Multiplier (EM) which represents the maximum eccentric contraction velocityas shown in Figure 1. This parameter ranges between 1.1 and 1.8 and varies from subject to subject.

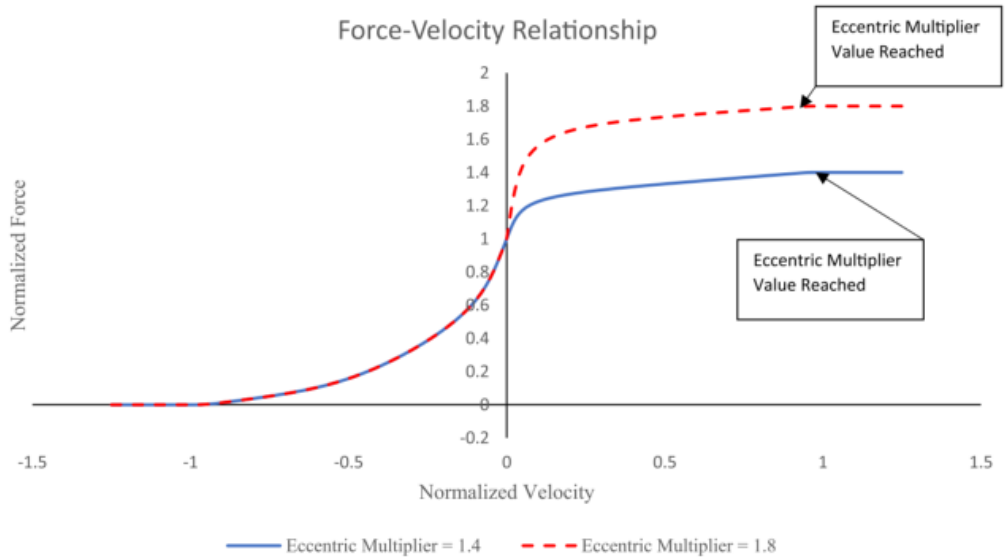

Figure 1: Force-velocity relationship. A higher eccentric multiplier results in higher forces during eccentric (lengthening) contractions. Positive velocity indicates lengthening and negative velocity indicates shortening.

Based on the eccentric multiplier, we consider three subjects: (i) Low force capacity (Sedentary or untrained): EM =1.2, (ii) Mid force capacity (Recreationally active): EM = 1.4, and (iii) High force capacity (Athletic): EM 1.8.

## 2.3 Impact characteristics

Based on the impact duration and impact velocity, we classified external forces into three categories: (i) Low-velocity impact (LVI): impact velocity $< 20 ms$ and impact durations $> 30 ms$ (e.g. automobile accidents), (ii) Intermediate velocity impact (IVI): (impact velocity $20 - 50 ms$ and impact durations $< 30 ms$) (e.g., sports impacts), (iii) High-velocity impact (HVI): impact velocity $> 50 ms$ and impact durations $< 5 ms$ (e.g., blasts).

## 2.4 Head Neck Injury Criteria:

Acceleration is measured in multiples of the acceleration of gravity (g), and time is measured in seconds. Proposed by the US National Highway Traffic Safety Administration (NHTSA) (Zhan et al., 2021) as shown in Eq. (5)

$$HIC = max \left[ \frac{1}{t_2-t_1} \int_{t_1}^{t_2} a(t) dt \right]^{2.5} (t_2 - t_1) \quad (5)$$

$HIC_{36}$ means that $t_2$ and $t_1$ not to be more than 36 ms apart. Based on animal and cadaver experiments. Suggested human tolerance for 50$^{th}$ percentile males: $HIC_{36} < 1000$ and $HIC_{15} < 700$. $HIC_{36} < 1000$ represents 18% probability of a severe head injury, 55% probability of serious injury, and 90% probability of a moderate head injury to the average adult (Mackay, 2007). In a study, concussions were found at $HIC = 250$ in most athletes (Viano, 2005; Viano & Pellman, 2005).

However, rotational acceleration is responsible for shear stress that damages brain tissue. HIC does not consider rotational acceleration of the head in its injury evaluation. Generalized Acceleration model Brain Injury Threshold (GAMBIT) model was developed by Newman (Newman, 1986) by combining translational and rotational components of head acceleration as shown in Eq. (6).

$$GAMBIT = \left[ \left(\frac{a(t)}{a_c}\right)^n + \left(\frac{\ddot{\phi}(t)}{\ddot{\phi}_c}\right)^m \right]^{\frac{1}{k}} \quad (6)$$

Where $a(t)$ and $\ddot{\phi}(t)$ denote the translational and rotational acceleration, respectively. $a_c$ and $\ddot{\phi}_c$ represent critical tolerance levels of those accelerations. $n, m,$ and $k$ are constants. A GAMBIT value of G=1 represents a 50% probability of serious brain injury (Feist et al., 2009).

In this study, we assess the effect of impact locations. The eccentric impacts create significant rotational accelerations. As HIC does not consider rotational acceleration, we used GAMBIT to assess the head injury risk. We used HIC values to validate our data in comparison with literature data.

As external impacts were horizontal and not compressive forces, we considered relative velocity and acceleration based neck injury criteria (NIC) (Boström & Kullgren, 2007). The Neck Injury Criteria (NIC) is shown in Eq. (7):

$$NIC(t) = H * a_{rel}(t) + v_{rel}(t)^2 \quad (7)$$

$a_{rel}(t), v_{rel}(t)$= Relative acceleration and velocity between the first thoracic vertebra (T1) and first cervical spine (C1), $H$ is the neck length (Boström & Kullgren, 2007). Predict injury considering the pressure gradient caused by a sudden change of the fluid flow inside the fluid compartment of the cervical spine. Bohman et al. (Bohman & Haland, 2000) suggested 15 $m^2/s^2$ human tolerance for AIS1 (minor injury). $N_{ij}$ is based on compressive force and moments on cervical spine and considers only flexion and extenion related injury. Therefore, we considered $NIC$ as we have lateral impacts, and it creates rotational movement along with flexion and extension.

## 2.5 Muscle-tendon strain

Muscle-tendon strain, expressed as a percentage increase of its current length compared to the length when it is at rest and developing no force. The resting length is called slack length. The strain of muscle-tendon at any given instant can be expressed in Eq. (8)

$$\epsilon^{MT} = \frac{l^{MT} - l_s^{MT}}{l_s^{MT}} \tag{8}$$

Here, $l^{MT}$ is muscle-tendon current length, and $l_s^{MT}$ is muscle-tendon slack length. In this study, we considered three hyoid muscles and three superficial multifidus muscles strain to assess the muscle injury probability. Three hyoid muscles: sternohyoid, sternothyroid, and omohyoid. The superficial multifidus muscles are: superficial multifidus(C5/6-C2), superficial multifidus(C6/7-C2), and superficial multifidus(T1-C4). For more than 10% strain, muscle-tendon begin to experience mechanical failure, and injury risk is high. However, some literature argues that muscle-tendon can sustain up to 15% of strain before it fails (Uchida & Delp, 2021).

## 2.6 Impact locations:

There are three different types of acceleration that affect the skull and the head: translational, rotational and angular acceleration, as shown in the Figure 2. For the translational type, also known as linear, the centre of gravity in the head will move in a straight line without rotation, and this usually results in focal injury, while in rotational acceleration, there is no movement of the center but the head will rotate around it resulting in diffuse shear strain. Lastly, the angular type is a combination of both rotational and linear properties, and the centre of gravity moves in an angular manner. The angular type is the most common type and is responsible for the bulk of TBI cases.

We consider seven impact locations. Two anterior impacts, two posterior impacts, and three lateral impacts. The two anterior impacts are anterior central (AC), and anterior eccentric (AE). The two posterior impacts are posterior central (PC) and posterior eccentric (PE). The lateral impacts are lateral central (LC), anterolateral eccentric (ALE), and posterolateral eccentric (PLE).

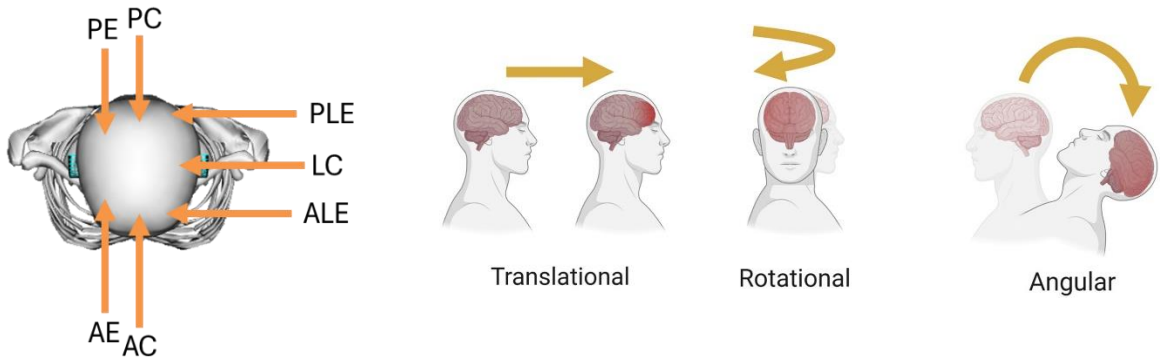

Figure 2 : Impact locations (AC = Anterior Concentric, AC= Anterior Eccentric, PC= Posterior Concentric, PE= Posterior Eccentric, LC= Lateral Concentric, PLE = Posterolateral Eccentric , ALE = Anterolateral Eccentric )

## 3 Experiment procedure and data collection:

### 3.1 OpenSim simulation Flow:

A modified head-neck musculoskeletal model was employed in OpenSim software. The model has Hill-type muscles and rate-dependent ligaments (Kuo et al., 2019). The model has total 72 muscle actuators. Previous studies verified the model using 10 subjects' data. Muscle activation data for neck stiffness were collected from the literature (Kuo et al., 2019). Three subjects, three impact types, and seven impact locations, a total of 63 cases, were performed. A Matlab script was used to run these 63 cases in OpenSim forward dynamics simulations. The workflow diagram is showin in Figure 3.

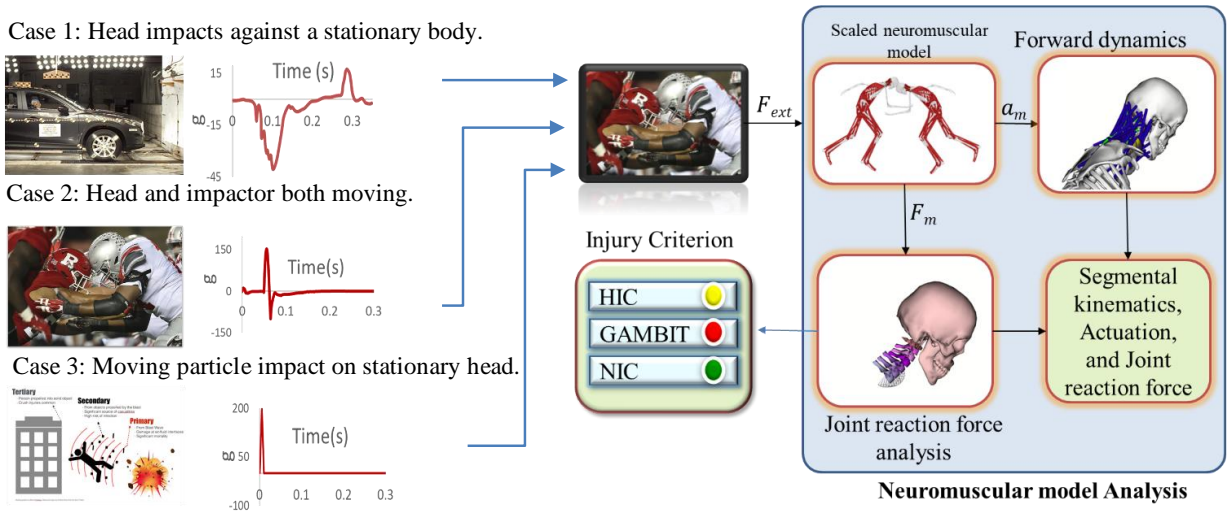

Figure 3: Musculoskeletal model simulation workflow

### 3.2 Data Collection

We collected three datasets to analyze the relationship of neck strength, impact locations, and impact characteristics on head and neck injury parameters. For the low-velocity impacts (LVI), we collected an automotive crash test from the NHTSA database. The test number was 11149 (National Highway Traffic Safety Administration, 2020). A frontal rigid barrier impact test was conducted on a 2020 Toyota Highlander SUV. The impact velocity of the vehicle was 56.35 km/h (35 mph). A $50^{th}$ percentile male Hybrid III ATD was placed in the driver sitting position. The impact duration was about 75ms, and the impact force was 1510N. For the intermediate-velocity impacts (IVI), we used extrapolated American Football data (Mortensen et al., 2018). The impact duration was 13 ms, and the maximum impact force was 1210 N. For the high-velocity impacts (HVI), we used simulated blast data (Kulkarni et al., 2014). The impact duration was 5ms and 3455N. The exact force and time duration data were used to validate the model.

# 4 Results

## 4.1 Accelerations:
The linear accelerations are presented in Figure 4 (a). The rotational accelerations are presented in Figure 4(b).

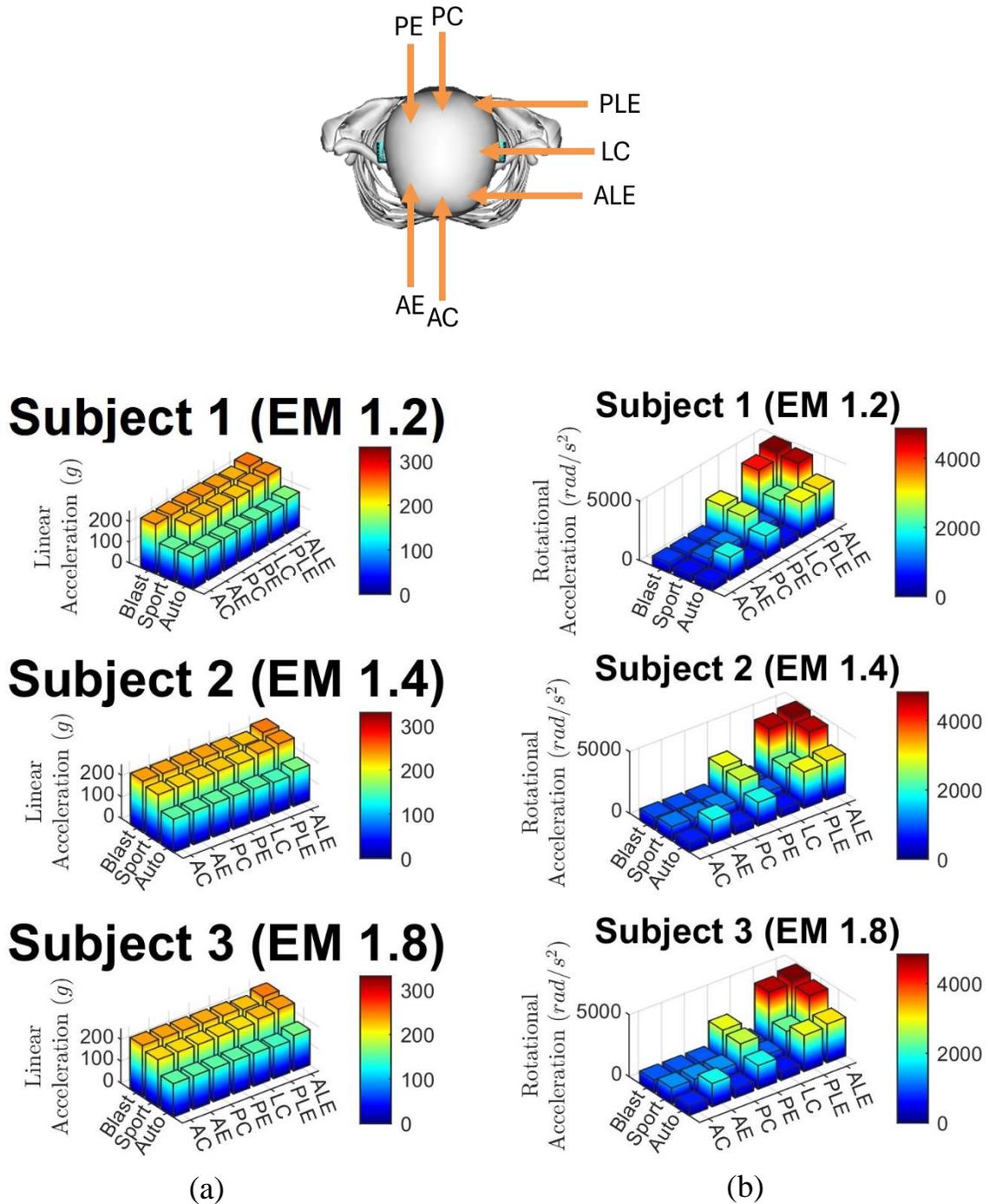

Figure 4: (a) Linear accelerations and (b) rotational accelerations for three subjects, three impact characteristics, and seven impact locations. (AC = Anterior Concentric, AC= Anterior Eccentric, PC=

Posterior Concentric, PE= Posterior Eccentric, LC= Lateral Concentric, PLE = Posterolateral Eccentric , ALE = Anterolateral Eccentric )

## 4.2 GAMBIT

The GAMBIT results are presented in Figure 5.

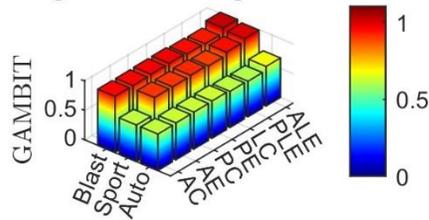

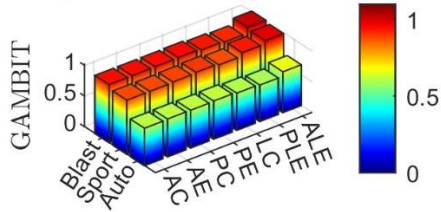

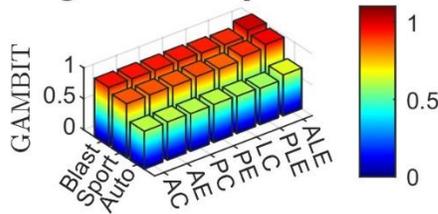

Figure 5: GAMBIT for three subjects, three impact characteristics, and seven impact locations.

## 4.3 Neck Force and Moment

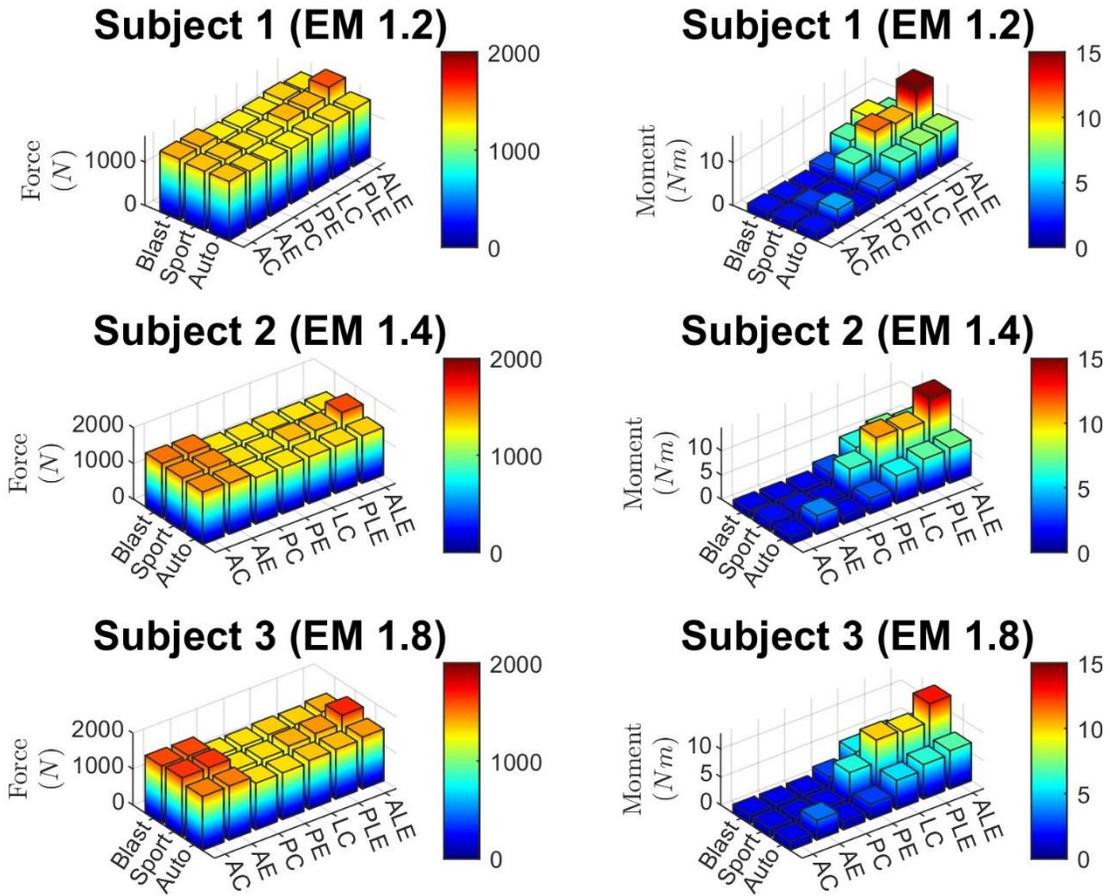

Figure 6: (a) Neck force, (b) neck moment for three subjects, three impact characteristics, and seven impact locations

## 4.4 NIC
The NIC data are presented in Figure 7.

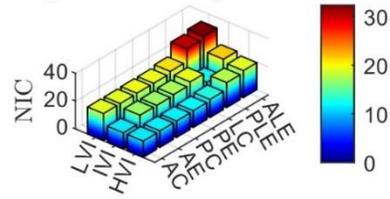

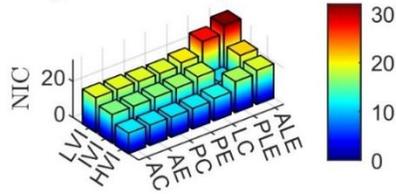

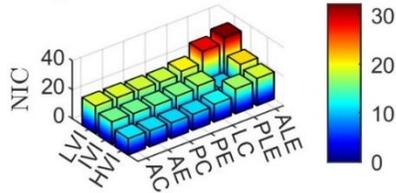

Figure 7: NIC for three subjects, three impact characteristics, and seven impact locations. (AC = Anterior Concentric, AC= Anterior Eccentric, PC= Posterior Concentric, PE= Posterior Eccentric, LC= Lateral Concentric, PLE = Posterolateral Eccentric , ALE = Anterolateral Eccentric )

## 4.5  Muscle strain
The six muscles' strains are presented in Figure 8.

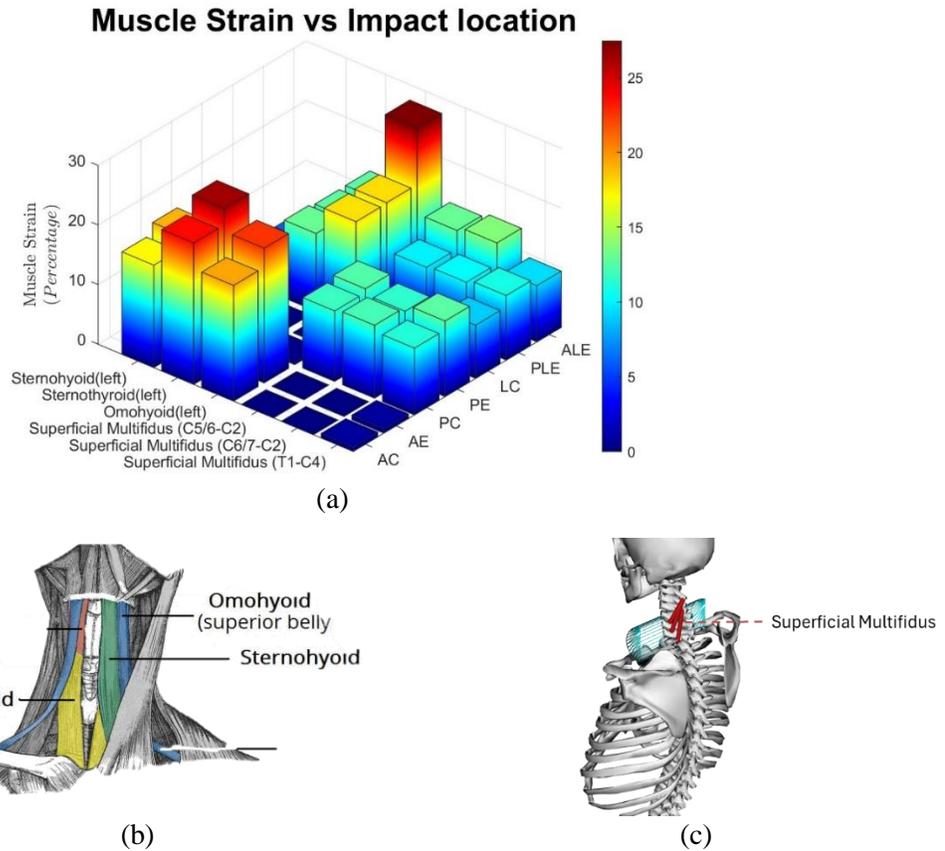

Figure 8: (a) Muscle Strain for three subjects, three impact characteristics, and seven impact locations. (b) hyoid muscles, (c) superficial multifidus muscles

## 5 Discussion

### 5.1 Effect of impact location:

Average maximum linear acceleration, rotational acceleration, GAMBIT, neck force, neck moment, and NIC are presented in igure 9 and Table I. It can be seen from 9 (a) that linear accelerations do not vary much based on impact direction and locations and stays within 194g to 218g. Eccentric impacts create higher linear acceleration than central impacts, however the increment is less than 10% percentage. The lateral impacts generate about 10% more linear accelerations than anterior and posterior impacts.

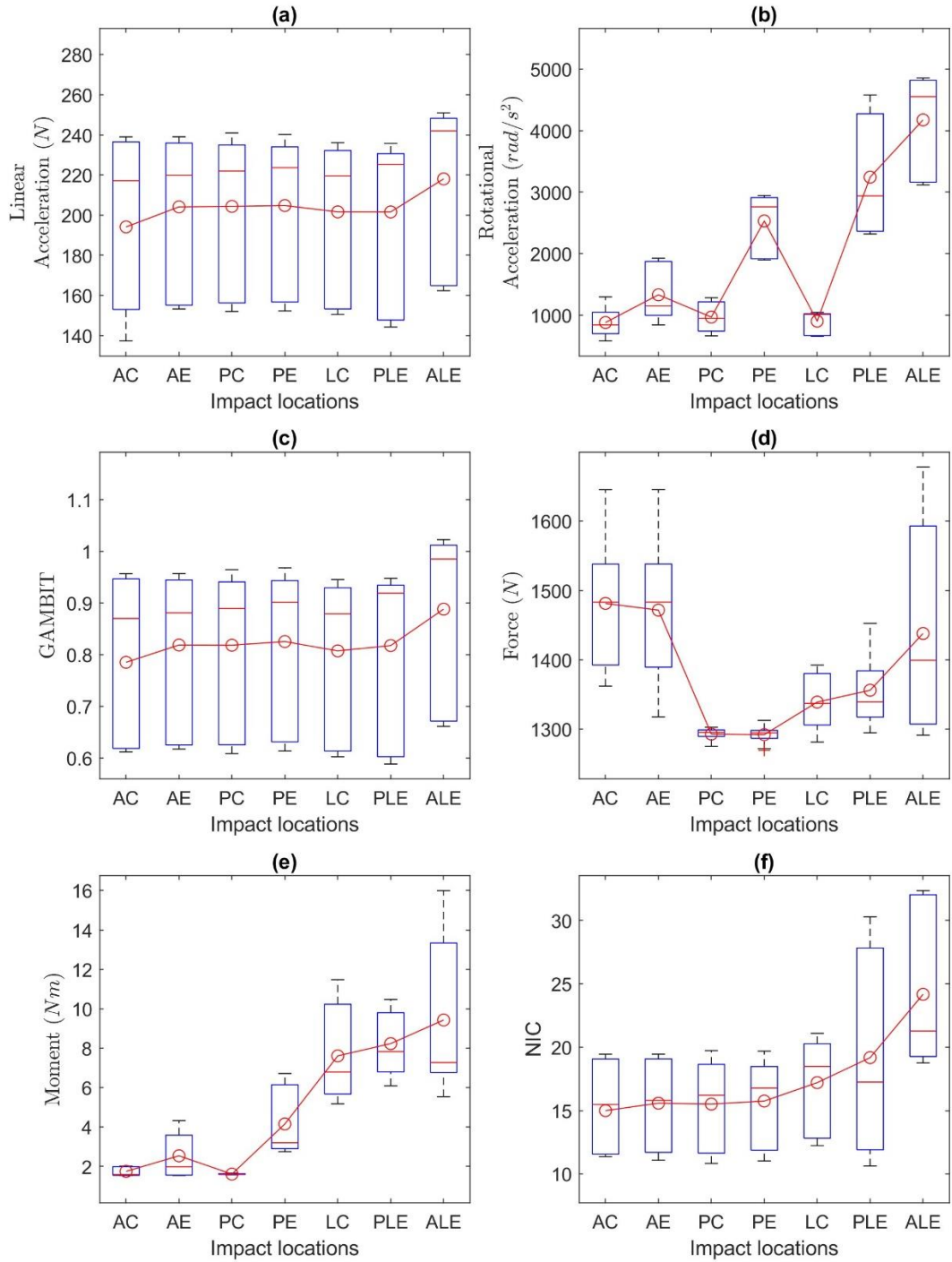

Figure 9: Effect of impact locations on (a) linear acceleration, (b) rotational acceleration, (c) GAMBIT, (d) neck force, (e) neck moment, and (f) NIC.

The rotational accelerations in Figure 9 (b) vary significantly based on impact directions and locations. The eccentric impacts generated about 1.51, 2.61, and 4.11 times more rotational acceleration than central impacts, for anterior, posterior, and lateral sides, respectively. Posterior and lateral impacts create 1.58 and 2.51 times more rotational acceleration than anterior impacts, respectively. The maximum average rotation acceleration is for anterolateral eccentric impact (4176 $rad/s^2$) which is 4.75 times more than the average anterior central impact (879 $rad/s^2$). The GAMBIT values in Figure 9(c) do not vary very much with the impact directions and locations similar to linear accelerations. It was within 0.79-0.83 for all impact locations except for posterolateral eccentric which was 0.89ss. In summary, impact locations have a significant effect on head and neck injury parameters, and anterolateral impact is the most risky impact location for both head and neck.

*Table 1: Effect of impact locations*

| Parameter | Anterior Central | Anterior eccentric | Posterior Central | Posterior Eccentric | Lateral Central | Posterolateral Eccentric | Anterolateral Eccentric |
|---|---|---|---|---|---|---|---|
| $a$ $(g)$ | 194 | 204 | 204 | 204 | 201 | 201 | 218 |
| $\alpha$ $(rad/s^2)$ | 879 | 1330 | 966 | 2531 | 901 | 3244 | 4176 |
| GAMBIT | 0.79 | 0.82 | 0.82 | 0.83 | 0.81 | 0.82 | 0.89 |
| $F_{neck}$ (N) | 1481 | 1471 | 1292 | 1291 | 1338 | 1356 | 1437 |
| $M_{neck}$ (Nm) | 1.75 | 2.54 | 1.61 | 4.16 | 7.61 | 8.23 | 9.43 |
| NIC | 14.9 | 15.59 | 15.52 | 15.75 | 17.21 | 19.18 | 24.17 |

The average neck forces in Figure 9 (d) do not vary much based on impact locations and directions and the variations of average neck forces were within 15 % (1290N -1481N). The maximum average neck forces was for the anterior central (1481N). Neck moments in Figure 9(e) vary significantly based on impact locations and directions. The highest average neck moment was for anterolateral eccentric (9.43Nm), which was 5.85 times more than the posterior central neck moment (1.61Nm). Also, eccentric impacts generate about 10-157% more neck moment than central impacts. NIC values in Figure 9 (f) were within 15-24 $m^2/s^2$. The lateral impacts values were higher than anterior and posterior impacts.

## 5.2 Effect of neck strength:

The effect of neck strength are presented in Figure 10 and Table 2. Average linear accelerations in Figure 10(a) do not vary much with neck muscle strength and were within $202g$ to $205g$. The average rotational accelerations in Figure (b) also do not vary much and stay between 1960 $rad/s^2$ to 2037 $rad/s^2$. The average GAMBIT in Figure 10(c) ranges from 0.81-0.83.

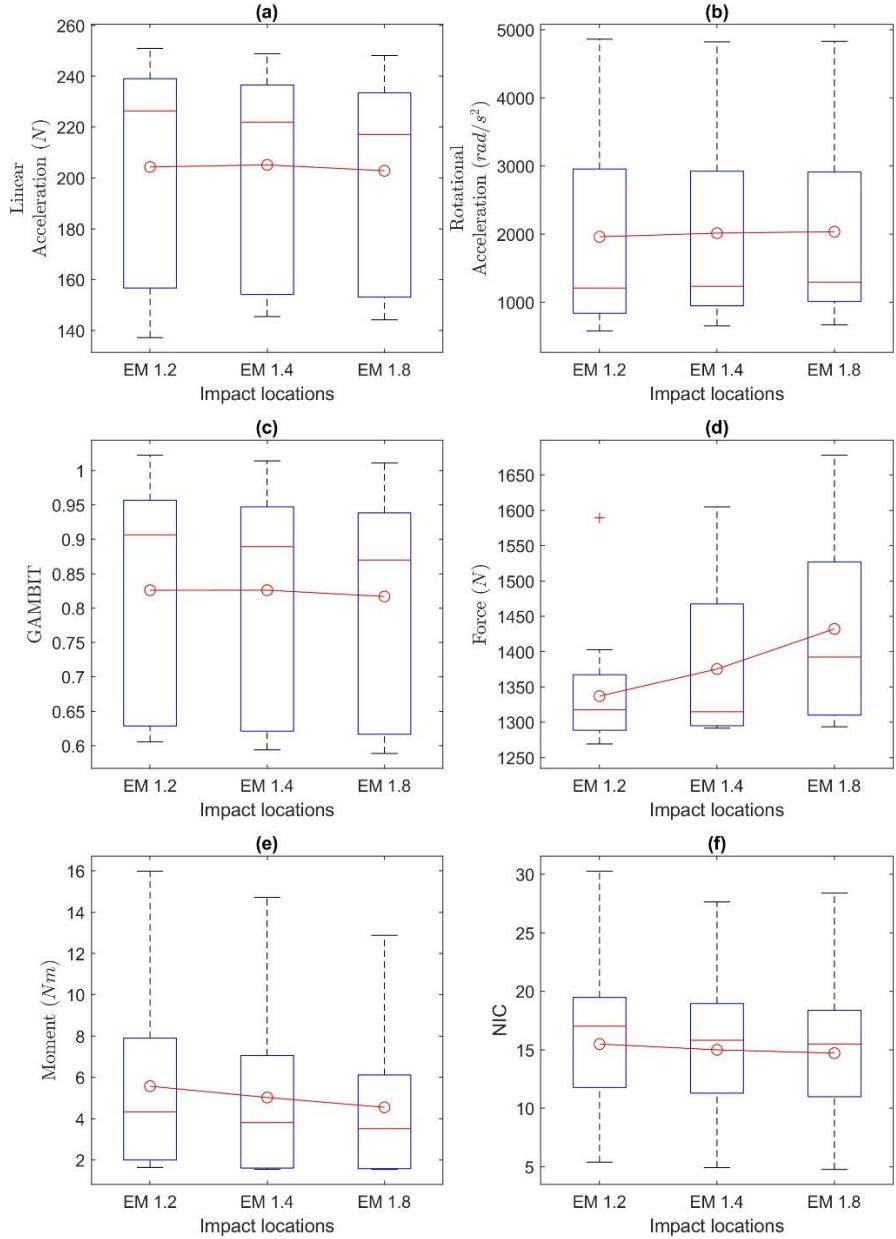

Figure 10: Effect of neck strength on (a) linear acceleration, (b) rotational acceleration, (c) GAMBIT, (d) neck force, (e) neck moment, and (f) NIC.

*Table 2: Effect of neck strength*

| Parameter | Low force capacity (EM 1.2) | Mid-force capacity (EM 1.4) | High force capacity (EM 1.8) |
|---|---|---|---|
| $a$ | 204 | 205 | 202 |
| $\alpha$ | 1962 | 2014 | 2036 |
| $GAMBIT$ | 0.82 | 0.82 | 0.81 |
| $F_{neck}$ | 1336 | 1375 | 1432 |
| $M_{neck}$ | 5.56 | 5.02 | 4.54 |
| $NIC$ | 15.4 | 14.98 | 14.7 |

The average neck force in Figure 10(d) ranges from 1336N to 1432N. The average neck moment in Figure 10(e) ranges from 4.5Nm to 5.6Nm. The average NIC value ranges from 14.7 to 15.5. In summary, average head and neck injury parameters do not vary more than 10% based on neck strength, which agrees with the literautre (Mortensen et al., 2020).

## 5.3 Effect of impact characteristics

The effect of impact characteristics are presented in Figure 11 and Table 3. The average linear accelerations in Figure 11(a) for IVI and HVI are 1.45 and 1.54 times higher than LVI, and range from $153g$ to $237g$. The maximum linear accelerations were 251g for HVI. The average rotational accelerations in Figure 11(b) for IVI and HVI are 1.18 and 1.35 times higher than LVI, and range from $1704\ rad/s^2$ to $2290\ rad/s^2$. The maximum rotational accelerations were $4859 rad/s^2$ for HVI. The average GAMBIT in Figure 11(c) for IVI and HVI were 1.44 and 1.54 times higher than LVI. The maximum GAMBIT was 1.02 for HVI.

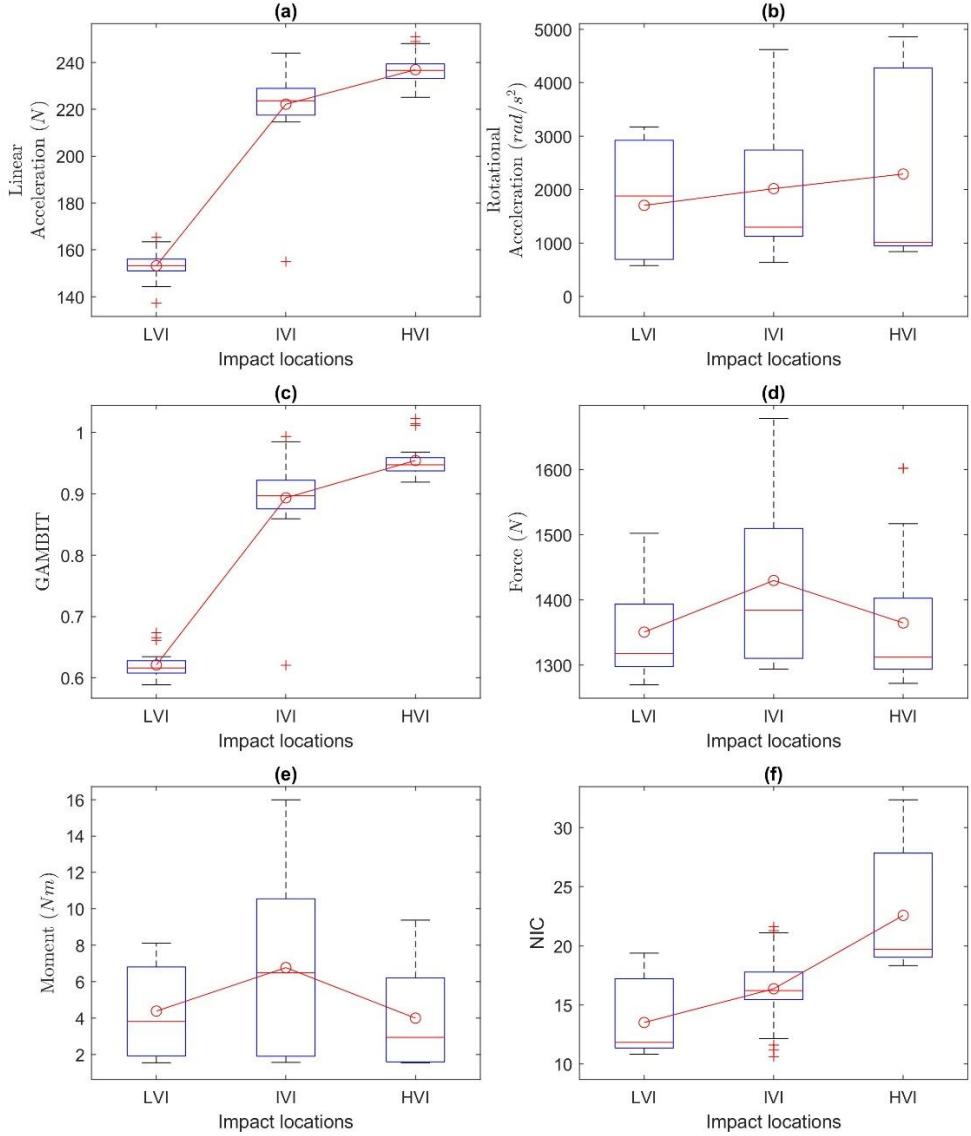

Figure 11: Effect of impact characteristics on (a) linear acceleration, (b) rotational acceleration, (c) GAMBIT, (d) neck force, (e) neck moment, and (f) NIC.

*Table 3: Effect of impact characteristics*

| Parameter | Low velocity impact (LVI) | Intermediate velocity impact (IVI) | High velocity impact (HVI) |
|---|---|---|---|
| $a$ | 154 | 223 | 237 |
| $\alpha$ | 1704 | 2018 | 2291 |
| $GAMBIT$ | 0.62 | 0.89 | 0.95 |
| $F_{neck}$ | 1350 | 1429 | 1394 |
| $M_{neck}$ | 4.38 | 6.77 | 4.29 |
| $NIC$ | 11.51 | 14.38 | 19.28 |

The neck forces and moments did not vary much based on the impact characteristics, as in Figure 11(d-e). The NIC in Figure 11(f) for IVI and HVI were 1.24 and 1.68 times more than LVI. The maximum NIC value was 30.28 for HVI. In summary, the risk probability of head and neck injuries is higher for HVI and IVI than LVI.

## 5.4 Validation:

The musculoskeletal model was validated in previous studies for IVI using NFL (Kuo et al., 2019; Mortensen et al., 2018, 2020) and rugby (Silvestros et al., 2024) data. However, it was not validated for LVI and HVI. To validate the model for LVI and HVI, we run the simulation using the literature data for LVI and HVI for all neck strength. Literature has only HIC value and max resultant linear acceleration that we compared with our result for the validation. However, neck strengths of the experimental subjects were not mentioned in the literature. So, we run the simulation for all neck strength. For LVI, the experimental data indicates that the subject's (ATD's) neck strength was between EM 1.2 and EM 1.4. For HVI, the experimental data indicates that subject's (ATD's) neck strength was between 1.4 and 1.8. The simulation results were consistent with the literature data.

*Table: Validation for LVI and HVI data*

|  | LVI | | HVI | |
| --- | --- | --- | --- | --- |
|  | HIC | Max resultant a (g) | $HIC_{15}$ | Max resultant a(g) |
| Experiment data | 292 | 57 | 710 | 122 |
| EM 1.2 | 303 | 61 | 740 | 143 |
| EM 1.4 | 284 | 55 | 723 | 139 |
| EM 1.8 | 273 | 53 | 703 | 133 |

Next, as we mentioned before, we run all the simulations for the same amplitude input force. However, the characteristics of the impact force pattern were taken from the literature. From our results, the rotational accelerations of the anterolateral eccentric and posterolateral eccentric are 4.7 and 3.7 times higher than the anterior central impacts, respectively. It is mentioned in literature that spin rotational acceleration (anterolateral eccentric or posterolateral eccentric) of the head is 3.5 times higher than whiplash acceleration (anterior impact), considering the head is a circle (Razzaghi, 2019). This data supports our acceleration data. The distance ratio from the front of the head to the center of mass of the head and from back of the skull to the center of mass of the skull is 0.52 and 0.48. That means the anterolateral impact will create more than the posterolateral impact. This also supports our simulation data.

## 5.5 Limitation

There are some limitations to this study. First, we ignored torso velocity and accelerations on the impact of head and neck injuries. We considered the torso as a fixed body. Secondly, we did not consider whole blast impacts on the head and neck injuries. We considered only the forces generated during a blast(HVI) impacting the head. This obviously does not perfectly reproduce the experimental environment and head impacts. Thirdly, we consider height, weight, and anthropometric data for all tests were same to consider the effects of neck strength, impacts locations, and impact characteristics on head and neck injuries. However, with different anthropometric and activation data, head-neck responses after an incident may vary. Also, we used the IVI muscle activation data for all cases. The activations data were collected from (Mortensen et al., 2018, 2020). Fourth, the subjects in this study are scaled as an adult. The anthropometric data of infants, children, and adults are different. Infants are not miniature adults. Infants and children have greater head-mass to body-mass ratio. Also, their head's center of gravity with respect to the neck is higher than that of adults (Burdi et al., 1969). All these factors may increase their TBI risk compared to adults for similar incidents (Eckner et al., 2014). The results on a scaled musculoskeletal model based on children may be different than this study and a scope for future researchers. Also, this study does not quantify the differences in male and female musculature.

# 6 Conclusion

In this study, we investigated the varying relationship of neck strength, impact locations, and impact characteristics with head and neck injury criteria. We employed a musculoskeletal model and used forward dynamics to run 63 cases. The results were validated using different literature data. It was found that impact locations have a significant effect on head and neck injury parameters, and anterolateral eccentric impact is the most risky impact location for both head and neck. We also found that average head and neck injury parameters do not vary more than 10% based on neck strength. Finally, the risk probability of head and neck injuries is higher for HVI and IVI than for LVI. These findings indicate that persons who face an anterolateral eccentric impact are at higher risk than the person who faces impact from other directions, and should seek medical attention. To our knowledge, this is the first study where head and neck injury parameters were assessed for multiple impact locations and for all three types of velocity impacts. Although the musculoskeletal-model-based framework has some limitations, it provides vital information about head and neck injuries for future researchers. The high biofidelity data from this study can be utilized as the input for tissue-level and molecular-level head and neck injury studies. That will enhance our understanding about the cause-and-effect of upper extremity organ-level injuries.